# Rotation free flow of noncompressible fluid through the rotating wheel of centrifugal pump

*Igor A. Tanski*
*Moscow, Russia*
*tanski.igor.arxiv@gmail.com*

*ABSTRACT*

The exact analytic solution is built for the plane flow of incompressible fluid through the wheel with profiles of blades approximating logarithmic spirals

**1. The geometry of the wheel and the flow through stationary cascade**

A. Busemann (ref. [1] ) built the exact solution for the flow through the rotating grid of zero thickness blades drawn by logarithmic spirals. Busemann constructed exact analytical solutions only for conformal mapping of grid and for the flow through the motionless grid. He expressed potential of the flow through rotating grid as Poisson integral and calculated the integral numerically.

T. S. Solomahova (ref. [2] ) repeated Busemann's calculations with computer use, which gave her possibility to increase accuracy of calculations, perform them for more broad parameters range and build graphs.

G. I. Majkapar (ref. [3]) constructed profile shape, which approximates logarithmic spirals. For this purpose he generalized Busemanns mapping. He considered stationary cascade only.

In this article we build exact solution for rotating cascade with Majkapar profile shape.

The basis of our considerations is following function, which transforms from one period of cascade on physical plane ($z$ plane) into unity circle on parameter plane ($t$ plane).

$$z = R \exp\left[ \frac{ln(1 + q/t) - a e^{2i\beta} ln(1 + qt)}{N} \right]; \tag{1}$$

where:

**R** - scale factor;

**N** - the number of rotor blades ;

**β** - constant angle between basic logarithmic blade and tangential direction;

**a** - dimensionless parameter ($0 \leq a \leq 1$), which determines the blades thickness, the case $a = 1$ corresponds to zero thickness blades;

**q** - dimensionless parameter ($0 \leq q \leq 1$), which determines the ratio of minimum and maximum radii, the case $q = 1$ corresponds to infinitely long blade.

Let us separate real and imaginary parts of (1). This gives expressions for Cartesian coordinates in the flow plane $z = x + iy$. In the parametric plane we use polar coordinates $t = \rho e^{i\phi}$. On the unity circle $\rho = 1$.

$$z = R \exp \left\{ \frac{1}{N} \left[ \tfrac{1}{2} \ln(1 + 2q \cos \phi + q^2) \, (1 - a \cos 2\beta) + \mathrm{arctg}\left( \frac{q \sin \phi}{1 + \cos \phi} \right) a \sin 2\beta - \right. \right. \tag{2}$$

$$\left. \left. -i \, (\tfrac{1}{2} \ln(1 + 2q \cos \phi + q^2) \, a \sin 2\beta + \mathrm{arctg}\left( \frac{q \sin \phi}{1 + \cos \phi} \right)(1 + a \cos 2\beta)) \right] \right\}.$$

Picture Fig. 1 illustrates the typical rotor form for parameter values $a = 0.85$ and $q = 0.9$. Picture Fig. 2 illustrates the case of very long profiles - $a = 0.85$ and $q = 0.999$. Picture Fig. 3 illustrates the case of very thick profiles - $a = 0.2$ and $q = 0.999$. On the picture Fig. 4 single profile for the normal case ($a = 0.85$ and $q = 0.9$) is pictured.

Let us determine the points on the profile, where radius takes its extreme values. For this purpose we differentiate real part of exponents argument in (2) on $\phi$ and equate the result to zero:

Let us denote:

$$\lambda = \frac{1 - ae^{2i\beta}}{N} \tag{3}$$

The radius square on the profile contour reads:

$$r^2 = z \, \bar{z} = R^2 \exp \left\{ Re(\lambda) \ln(1 + 2q \cos \phi + q^2) - 2 \, Im(\lambda) \, \mathrm{arctg}\left( \frac{q \sin \phi}{1 + \cos \phi} \right) \right\} \tag{4}$$

After differentiation and simplifications we get the following equation for the angle $\phi$ value in extreme points:

$$Re(\lambda) \sin(\phi) + Im(\lambda) \, (q + \cos \phi) = 0. \tag{5}$$

This equation we can write in another form:

$$Im(\lambda(q + t)) = 0. \tag{6}$$

The equation (6) has 2 solutions:

$$\phi_1 = -Arg(\lambda) + \pi + arcsin(q \sin(Arg(\lambda))); \tag{7}$$

$$\phi_2 = -Arg(\lambda) - arcsin(q \sin(Arg(\lambda))). \tag{8}$$

First of them corresponds to minimum radius, the second - to maximum.

Substitute (3) to (5) and get another form of equation:

$$\frac{q + \cos(\phi_2)}{\sin(\phi_2)} = \frac{1 - a\cos(2\beta)}{a\sin(2\beta)}. \tag{9}$$

The complex potential of the flow through the cascade is a sum of two components. The first component is

$$w_1 = \frac{Q - i\Gamma_1}{2\pi} \ln(t+q) - \frac{Q + i\Gamma_2}{2\pi} \ln(t) + \frac{Q + i\Gamma_1}{2\pi} \ln(t + 1/q). \tag{10}$$

where
**Q** - the volume flow rate through the pump per one blade;
$\Gamma_1$ - the inlet circulation per one blade;
$\Gamma_2$ - the outlet circulation per one blade.

The first component $w_1$ is the potential of flow through stationary cascade. The boundary conditions in the infinity before and after the cascade are satisfied by this solution, but the boundary conditions on the blades surface corresponds to zero rotation speed.

The second component $w_2$ is the potential of flow through rotating cascade with zero boundary conditions in the infinity before and after the cascade. Busemann [3] called it - displacement flow; other authors refer to its rotating cells as "relative eddies".

## 2. Displacement flow

For the displacement flow the boundary conditions on the blades surface are (ref. [1], chapter 4, section 72):

$$\psi_2 = -\frac{1}{2}\omega r^2. \tag{11}$$

where $\psi$ is the flux function (imaginary part of the complex potential).

Thus, as to find complex potential for the displacement flow we have to build analytic function by its imaginary part values on the unity circle. We solve this problem using the Fourier series method.

On the boundary of the unity circle of the parametric plane identity $\bar{t} = 1/t$ holds. Therefore we can write (4) in the form:

$$r^2 = R^2 \exp[\lambda \ln(1 + qt) + \bar{\lambda} \ln(1 + q/t)]. \tag{12}$$

This mean, that $r^2$ boundary values coincide with boundary values of analytic function, which has singularities in the points $t = -q$ and $t = 0$ only. We shall use this fact to calculate integrals using Cauchy formula.

So as to get coefficients of Fourier series for $r^2$ we need to calculate following integrals:

$$A_m = \frac{1}{2\pi R^2} \int_{-\pi}^{\pi} r^2(\phi) e^{im\phi} d\phi. \tag{13}$$

Let us calculate contour integral:

$$I = \int_C \exp[\lambda \ln(1 + qt) + \bar{\lambda} \ln(1 + q/t)] \, t^{m-1} dt, \tag{14}$$

The integration path $C$ is presented on the picture Fig. 5. Contour $D$, which we use in the next equation, is the "2-3-4-5-6" circle arc.

Integral $I$ is zero, because its integrand is analytic function without singularities inside of the contour $C$. Let us decompose the $I$ integral.

$$I = \int_D (1 + qt)^{\lambda} (t + q)^{\bar{\lambda}} (-t)^{m-\bar{\lambda}-1} dt (e^{-i\pi(m-\bar{\lambda}-1)} - e^{i\pi(m-\bar{\lambda}-1)}) - 2\pi i A_m = 0. \tag{15}$$

Integrals across segments (1-2) and (6-7) are mutually cancelled, because integrand's many-valued powers contain fractional term $\bar{\lambda}$ with different signs. Therefore the integrand returns to the initial value when we go around of both branch point.

The integral along the real axis segment in (14) can be expressed through hypergeometric function. Finally we get the following expression for Fourier series coefficients

$$A_m = \frac{q^m \Gamma(\bar{\lambda} + 1)}{\Gamma(1 - m + \bar{\lambda})\Gamma(m + 1)} F(-\lambda, m - \bar{\lambda}; m + 1; q^2). \tag{16}$$

Separate real part of complex series and get

$$r^2 = R^2[A_0 + 2 \sum_{m=1}^{\infty} Re(A_m) \cos(m\phi) + Im(A_m) \sin(m\phi)] = R^2 \text{Re}[\bar{A}_0 + 2 \sum_{m=1}^{\infty} \bar{A}_m e^{im\phi}]. \tag{17}$$

(16) and boundary condition (10) for flux function imply the following Taylor series for the displacement flow potential

$$w_2 = -i\omega R^2 \, [\tfrac{1}{2} \bar{A}_0 + \sum_{m=1}^{\infty} \bar{A}_m t^m]. \tag{18}$$

Sum expressions (9) and (17) for two components and get full expression for complex potential

$$w = \frac{Q - i\Gamma_1}{2\pi} \ln(t+q) - \frac{Q + i\Gamma_2}{2\pi} \ln(t) + \frac{Q + i\Gamma_1}{2\pi} \ln(t + 1/q) - i\omega R^2 \, [\tfrac{1}{2} \bar{A}_0 + \sum_{m=1}^{\infty} \bar{A}_m t^m]. \tag{19}$$

## 3. Expression for the circulation

The inlet and outlet circulations are connected by a certain relation. For profiles with a cusp point Joukowski - Chaplygin postulate delivers such a relation. According to this postulate the rear critical point must coincide with the cusp point. Profiles under consideration have dont have any cusp point.

Let us suppose that the rear critical point coincide with the maximum radius point. For Busemanns logarithmic profiles of zero thickness this condition coincides with Joukowski - Chaplygin postulate.

The following condition holds for the critical point

$$\frac{dw}{dt} = 0, \quad t = t_c. \tag{20}$$

We use this condition so as to determine the outlet circulation:

$$\Gamma_2 = \frac{2\sin(\phi_c)\, Q + \Gamma_1(q - 1/q)}{2\cos(\phi_c) + q + 1/q} - \omega\, 2\pi R^2 \sum_{m=1}^{\infty} m\, \bar{A}_m t_c^m \tag{21}$$

where $t_c$ and $\phi_c$ are $t$ and $\phi$ values in the rear critical point.

## 4. Simplifications

Expressions (15), (18) and (20) give full solution of the problem. In particular, from (20) we can find the head of the pump (see below).

Expression (2) is inconvenient for calculations, because its last term is infinite series and its components are also infinite series according to (15). Besides that by $q = 1$ infinite series (20) diverges. Its summation is still possible using special methods (Cesaro for example).

Fortunately it is possible to write (20) in another form, using known properties of hypergeometric function and equation (6) for rear critical point. Let us denote

$$P(t) = \sum_{m=1}^{\infty} \bar{A}_m t^m. \tag{22}$$

We use the following Gauss relations for contiguous hypergeometric functions [Stegun]

$$c\,(1-z)\,F(a,b;c;z) - c\,F(a,b-1;c;z) + (c-a)\,z\,F(a,b;c+1;z) = 0; \tag{23}$$

$$(c-b-1)\,F(a,b;c;z) + c\,F(a,b+1;c;z) - (c-1)\,z\,F(a,b;c-1;z) = 0; \tag{24}$$

Using these identities, we can get finite differences equation for $A_m$ (we omit details):

$$\bar{A}_m\,(m-\lambda) + \bar{A}_{m-1}\,[(m+1)\,(q+1/q) + q\,(\lambda-\bar{\lambda})] + \bar{A}_{m+2}\,(m+2+\bar{\lambda}) = 0. \tag{25}$$

Equation (24) results in the following differential equation for $P$:

$$t[1 + (q+1/q)/t + 1/t^2]P' + [-\lambda + q(\bar{\lambda}-\lambda) + \bar{\lambda}/t^2]P + \tag{26}$$
$$+\bar{A}_1[-(q+1/q) - q(\bar{\lambda}-\lambda) - (1+\bar{\lambda})/t] - \bar{A}_2(2+\bar{\lambda}) = 0.$$

Note, that for rear critical point coefficient by $P$ in (25) is zero. This follows from (6):

$$(\lambda t - \bar{\lambda}/t) + q(\lambda-\bar{\lambda}) = 2i\,\text{Im}(\lambda t) + 2qi\,\text{Im}(\lambda) = 2i\,\text{Im}(\lambda t + q\lambda) = 0. \tag{27}$$

We need expression for $tP'$ in rear critical point to simplify (20). Using (26), we can write as follows:

$$tP'|_{t_c} = \frac{\bar{A}_1[(q+1/q) + q(\bar{\lambda}-\lambda) + (1+\bar{\lambda})/t_c] + \bar{A}_2(2+\bar{\lambda})}{1 + (q+1/q)/t_c + 1/t_c^2}. \tag{28}$$

Apply finite differences equation (24) for $m=0$ to RHS of expression (27) and get:

$$tP'|_{t_c} = \frac{\lambda t_c \bar{A}_0 + \bar{A}_1(1+\bar{\lambda})}{t_c + 1/t_c + q + 1/q}. \tag{29}$$

This expression does not contain infinite series for $P'$. But it still contains infinite series for hypergeometric functions. To make this clear, let us express $\bar{A}_0$ and $\bar{A}_1$ in (28) through hypergeometric functions values.

According to (15), we have the following expressions for $A_0$ and $A_1$:

$$\bar{A}_0 = F(-\bar{\lambda}, -\lambda; 1; q^2); \tag{30}$$

$$\bar{A}_1 = q\lambda F(-\bar{\lambda}, 1-\lambda; 2; q^2). \tag{31}$$

We shall use the following special case of Gauss identity (23) by $a=-\bar{\lambda}$, $b=-\lambda$, $c=2$.

$$(1 + \bar{\lambda})F(-\bar{\lambda}, 1 - \lambda; 2; z) - F(-\bar{\lambda}, -\lambda; 1; z) = \tag{32}$$

$$= [(1 + \lambda)(1 + \bar{\lambda})F(-\bar{\lambda}, -\lambda; 2; z) - (1 + \lambda + \bar{\lambda})F(-\bar{\lambda}, -\lambda; 1; z)]/\lambda,$$

Substitute (29) and (30) to (31) and get

$$(1 + \bar{\lambda})\bar{A}_1 - \lambda q \bar{A}_0 = q[(1 + \lambda)(1 + \bar{\lambda})F(-\bar{\lambda}, -\lambda; 2; q^2) - (1 + \lambda + \bar{\lambda})F(-\bar{\lambda}, -\lambda; 1; q^2)]. \tag{33}$$

This gives following expression for numerator of (28):

$$\lambda t_c \bar{A}_0 + (1 + \bar{\lambda})\bar{A}_1 = \lambda(t_c + q)F(-\bar{\lambda}, -\lambda; 1; q^2) + \tag{34}$$

$$+q[(1 + \lambda)(1 + \bar{\lambda})F(-\bar{\lambda}, -\lambda; 2; q^2) - (1 + \lambda + \bar{\lambda})F(-\bar{\lambda}, -\lambda; 1; q^2)]$$

Using these identities, we can write expression (20) for the outlet circulation as follows:

$$\Gamma_2 = \frac{2\sin(\phi_2)Q + (q - 1/q)\Gamma_1}{2\cos(\phi_2) + q + 1/q} - \tag{35}$$

$$-\omega 2\pi R^2 \frac{\lambda(t + q)F(-\bar{\lambda}, -\lambda; 1; q^2) + q((1 + \lambda)(1 + \bar{\lambda})F(-\bar{\lambda}, -\lambda; 2; q^2) - (1 + \lambda + \bar{\lambda})F(-\bar{\lambda}, -\lambda; 1; q^2))}{2\cos(\phi_2) + q + 1/q}.$$

(34) is the final form of outlet circulation expression.

## 5. Dimensionless coefficients

Recall, that we considered the volume flow rate through the pump, inlet and outlet circulations per one blade. We can get the full flow rate and circulations using the following obvious expressions:

$$Q_N = NQ; \quad \Gamma_{1N} = N\Gamma_1; \quad \Gamma_{2N} = N\Gamma_2; \tag{36}$$

Let us write expression for full outlet circulation in following standard form:

$$\Gamma_{2N} = -\mu\Gamma_{1N} + (1 - \mu)ctg(\alpha_0)Q_N - \omega 2\pi r_2^2 Bu, \tag{37}$$

where dimensionless coefficients are:

$\mu$ - transparency coefficient;

$\alpha_0$ - inlet angle for flow with zero circulation flow;

**Bu** - Busemann's head rise coefficient.

Put together (36) and (34) and get expressions for dimensionless coefficients:

$$\mu = \frac{1/q - q}{2\cos(\phi_2) + q + 1/q}; \tag{38}$$

$$\alpha_0 = arctg\left(\frac{q + \cos(\phi_2)}{\sin(\phi_2)}\right); \qquad (39)$$

$$Bu = \frac{NR^2}{r_2^2} \frac{\lambda(t_c + q)F(-\bar{\lambda}, -\lambda; 1; q^2)}{2\cos(\phi_2) + q + 1/q} + \qquad (40)$$

$$+ \frac{NR^2}{r_2^2} q \frac{(1+\lambda)(1+\bar{\lambda})F(-\bar{\lambda}, -\lambda; 2; q^2) - (1 + \lambda + \bar{\lambda})F(-\bar{\lambda}, -\lambda; 1; q^2)}{2\cos(\phi_2) + q + 1/q}.$$

According to (8') we have also

$$\alpha_0 = arctg\left(\frac{1 - a\cos(2\beta)}{a\sin(2\beta)}\right). \qquad (41)$$

This means, that in fact $\alpha_0$ does not depend on $q$.

We can solve (41) for $\beta$ and get:

$$\beta = \frac{1}{2}\left[\alpha_0 + arccos\left(\frac{\cos(\alpha_0)}{a}\right)\right]. \qquad (42)$$

## 6. $q = 1$ - long blades case

Important special case is $q = 1$. This case corresponds to $r1/r2 = 0$. In fact even by $r1/r2 = 0.6$ $q$ is very close to unity - therefore we can use for Busemann coefficient simplified formula, which does not contain $q$.

In this case we can use known Gauss expression for hypergeometric function with unity argument

$$F(a, b; c; 1) = \frac{\Gamma(c)\Gamma(c-a-b)}{\Gamma(c-a)\Gamma(c-b)}, \qquad (43)$$

where $\Gamma$ is Eulers gamma function.

(39) reads now:

$$Bu = \frac{NR^2}{r_2^2} \frac{\Gamma(\lambda + \bar{\lambda})}{\Gamma(\lambda)\Gamma(\bar{\lambda})}, \quad by \ q = 1. \qquad (44)$$

## 7. Thin radial blades case

Another important special case is $a = 1, \beta = \pi/2, 0 \leq q \leq 1$. In this case the dependencies for $\mu$ coefficient are very simple.

In this case according to (3)

$$\lambda = \frac{2}{N}, \tag{45}$$

and according to (8)

$$\phi_2 = 0. \tag{46}$$

Then (39) gives

$$\alpha_0 = \frac{\pi}{2}. \tag{47}$$

(4) provides following expressions for $r_{min}$ and $r_{max}$:

$$r_{min} = (1-q)^{2/N}; \tag{48}$$

$$r_{min} = (1+q)^{2/N}, \tag{49}$$

or

$$\frac{r_{min}}{r_{max}} = \left(\frac{1-q}{1+q}\right)^{2/N}. \tag{50}$$

(37) reads now

$$\mu = \frac{1-q}{1+q}. \tag{51}$$

We can solve (50) for $q$ and substitute to (51). This gives dependency of $\mu$ from radii ratio.

$$\mu = \left(\frac{r_{min}}{r_{max}}\right)^{N/2}. \tag{52}$$

## 8. Expression for the head rise of the wheel

Let us find the pumps head rise. According to Eulers formula, it is equal to

$$H = \omega \, (r_2 V_{\phi 2} - r_1 V_{\phi 1}), \tag{53}$$

where

$V_{\phi 1}$ - input tangential velocity;

$V_{\phi 2}$ - output tangential velocity.

We can express the head rise through inlet and outlet circulations:

$$H = -\frac{\omega}{2\pi}(\Gamma_{1N} + \Gamma_{2N}). \tag{54}$$

Substitute expression (36) for $\Gamma_2$ and get

$$H = -\frac{\omega}{2\pi}[(1-\mu)\Gamma_{1N} + (1-\mu)\,ctg(\alpha_0)\,Q_N - \omega\,2\pi r_2^2\,Bu]. \tag{55}$$

Let us introduce two new variables, which are more closely related to "physical" pump:

$\beta_1$ - inlet flow angle;

$\psi$ - ratio of tangential velocity of rotating wheel to average radial outlet velocity.

$\beta_1$ is determined by inlet helix construction. $\psi$ determines operating conditions of the pump.

We can express these variables as follows:

$$ctg(\beta_1) = -\frac{\Gamma_{1N}}{Q_N}; \tag{56}$$

$$\psi = \omega\,\frac{2\pi r_2^2}{Q_N}. \tag{57}$$

In these new variables expression (44) for the head rise is:

$$H = -\frac{\omega Q_N}{2\pi}[(1-\mu)(ctg(\alpha_0) - ctg(\alpha_1)) - \psi Bu]. \tag{58}$$

In the limit $N \to \infty$ we have $\mu \to 0$, $Bu \to 1$ and $H \to H_\infty$. Expression for $H_\infty$ is

$$H_\infty = -\frac{\omega Q_N}{2\pi}(ctg(\alpha_0) - ctg(\alpha_1) - \psi). \tag{59}$$

The ratio of real head rise $H$ to $H_\infty$ is

$$k_z = \frac{H}{H_\infty} = \frac{(1-\mu)(ctg(\alpha_0) - ctg(\alpha_1)) - \psi Bu}{ctg(\alpha_0) - ctg(\alpha_1) - \psi} \tag{60}$$

## 9. Numerical results

We use these formulae to calculate cascade characteristics.

The geometry of profile is determined by values of 4 parameters. We choose for this purpose following parameters: number of blades $N$, coefficient of thickness $\sigma$, inlet angle for flow with zero circulation flow $\alpha_0$ and radii ratio $r1/r2$. The coefficient of thickness $\sigma$ is calculated for average radius:

$$\rho = \sqrt{r_1 r_2} \tag{61}$$

The expression for $\sigma$ is

$$\sigma = \frac{2\pi - N\Delta\phi}{2\pi}. \tag{62}$$

where $\Delta\phi$ is the angle between two vectors going from zero to points with $\rho$ radius, laying on the blades profile.

We selected values of $N$, $\sigma$, $\alpha_0$, $r1/r2$, calculated corresponding parameters of conformal mapping (1) - $a$, $\beta$, $q$, $R$. After that calculated transparency coefficient $\mu$ and Busemann's head rise coefficient $Bu$.

In this way we obtained dependencies of transparency coefficient $\mu$ and Busemann's head rise coefficient $Bu$ from geometric parameters of the profile.

On the graph Fig. 6 dependency of Busemann's head rise coefficient $Bu$ from the number of blades is presented for the special case $a = 1, q = 1, \beta = \pi/2$. This is a case of thin radial blades with $r_1/r_2 = 0$. Obviously, $\mu = 0$ for this case.

Pictures Figs. 7-8 traits a slightly more general case. The blades are still thin and radial, but $r_1/r_2$ varies. The picture Fig. 7 gives dependency graph for transparency coefficient $\mu$, Fig. 8 - for Busemann's head rise coefficient $Bu$.

On the graphs Fig. 9-14 dependencies of $\mu$ and $Bu$ from inlet angle for flow with zero circulation flow $\alpha_0$ and coefficient of thickness $\sigma$ by fixed value of radii ratio $r1/r2 = 0.6$ are presented.

We see on the graphs, that coefficient of thickness $\sigma$ strongly effects the $\mu$ and $Bu$. By approximated calculations the effect of $\sigma$ sometimes is neglected. The graphs Fig.9-14 give possibility to evaluate inaccuracy of this approximation.

**DISCUSSION**

The exact analytic solution is built for the plane flow of inviscid incompressible fluid in the wheel with profiles of blades approximating logarithmic spirals.

______________________________

**REFERENCES**

[1] A. Busemann, Das Vorderhöchenverhaltnis radialer Kreiselpumpen mit logarithmischspiraligen Schaufeln, ZAMM, 1928, Bd 8, S. 372

[2] T. S. Solomahova, Raschet aerodinamicheskih harakteristik vraschayuschihsya krugovyh reshetok profilej, ocherchennyh po logarifmicheskim spiralyam, Promyshlennaya aerodinamika, 1966, Vol. 28, pp. 33 - 59

[3] G. I. Majkapar, Raschet krugovyh reshetok, Promyshlennaya aerodinamika, 1966, Vol. 28, pp. 3 - 17

**APPENDIX**

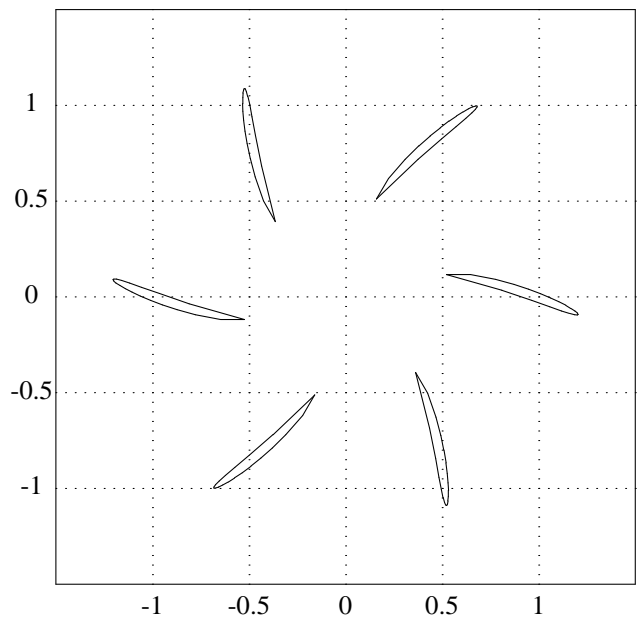

Fig 1. Typical grid: a = 0.85, q = 0.9

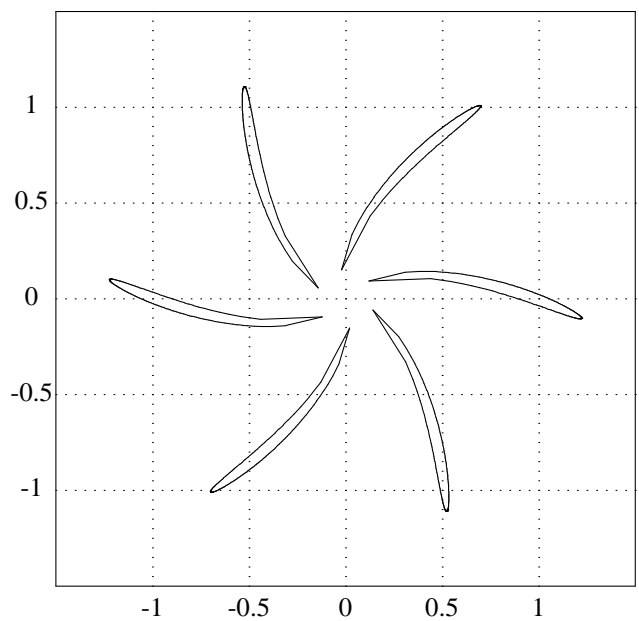

Fig 2. Very long profiles: a = 0.85, q = 0.999

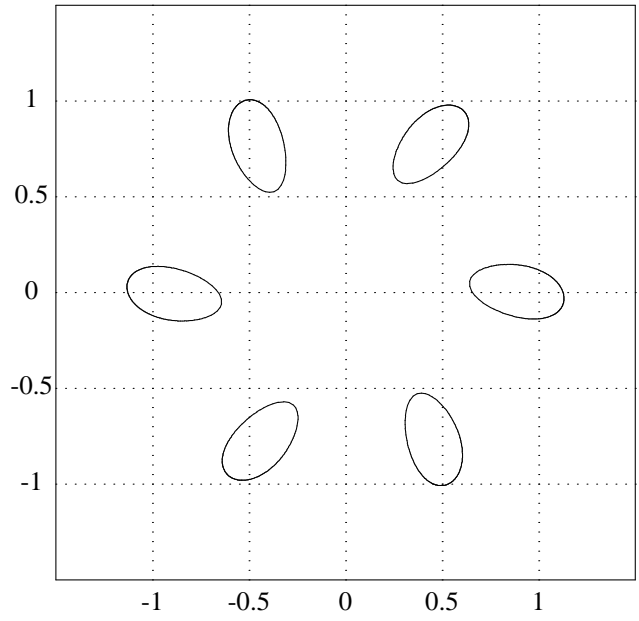

Fig 3. Very thick profiles: a = 0.2, q = 0.9

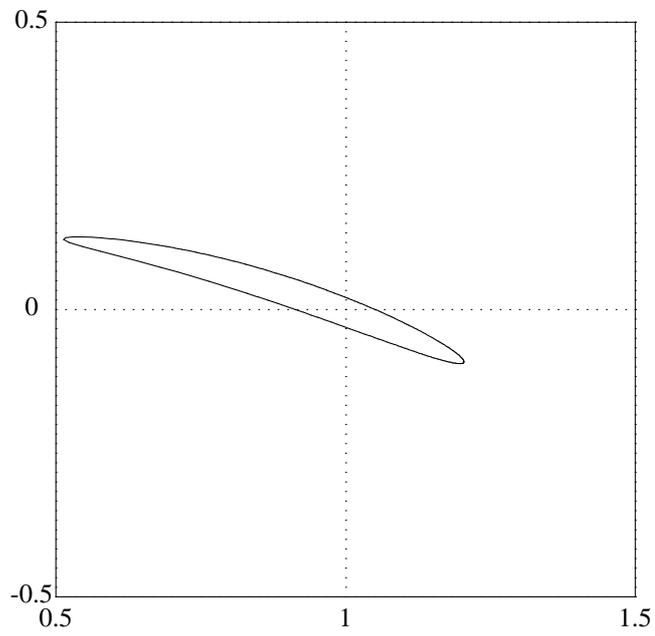

Fig 4. Single profile: a = 0.85, q = 0.9

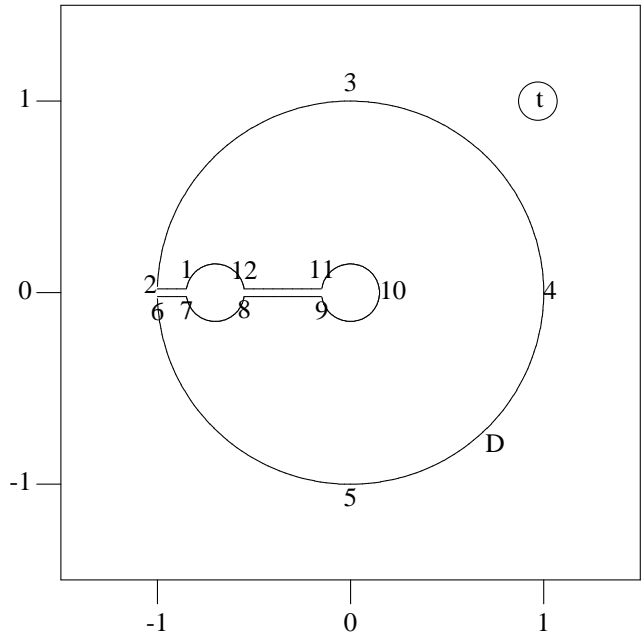

Fig 5. The integration path C

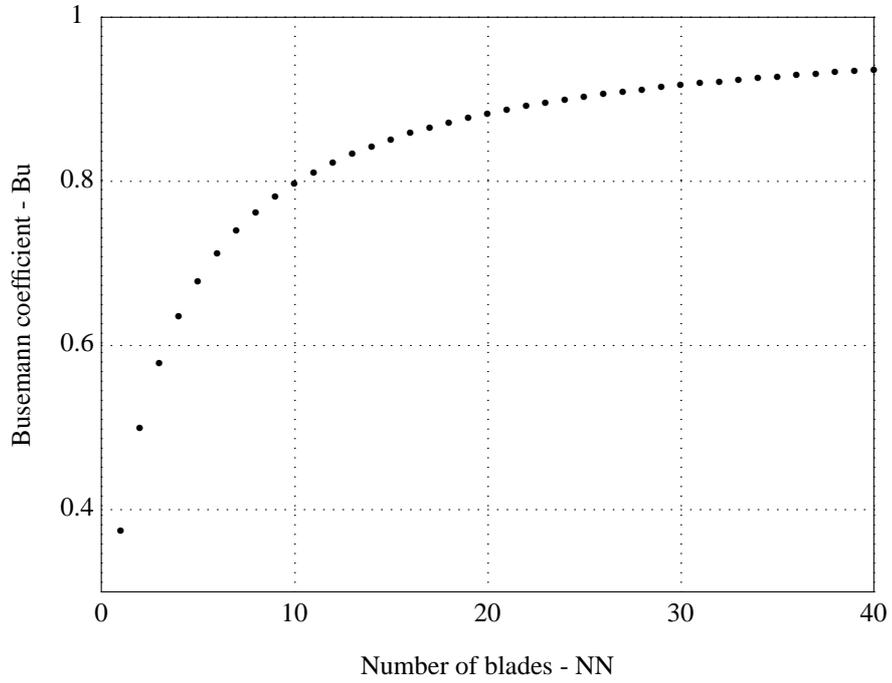

Fig 6. Thin long straight blades a=1, q=1, $\beta = \pi/2$

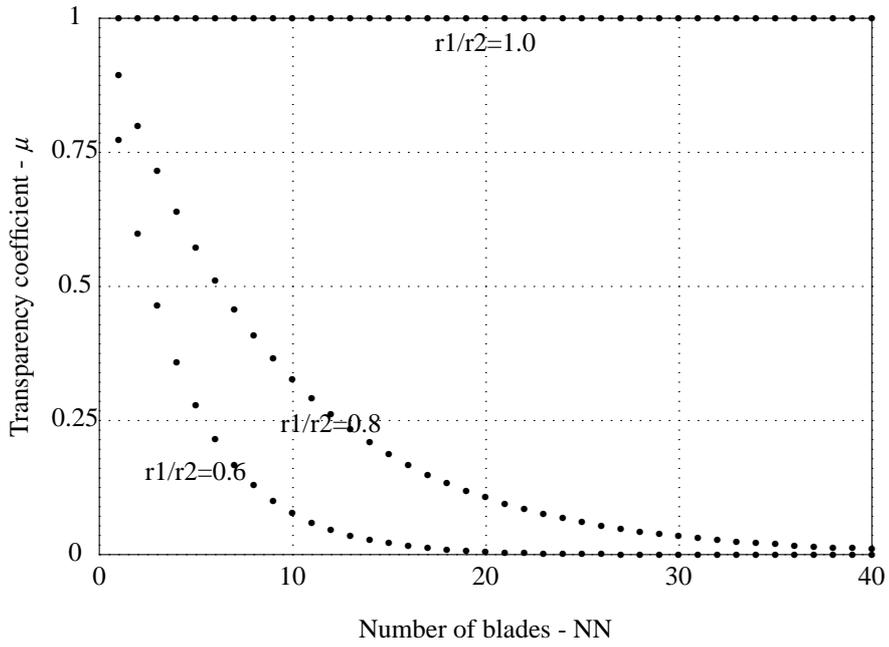

Fig 7. Thin straight blades $a = 1, 0 \leq q \leq 1, \beta = \pi/2$

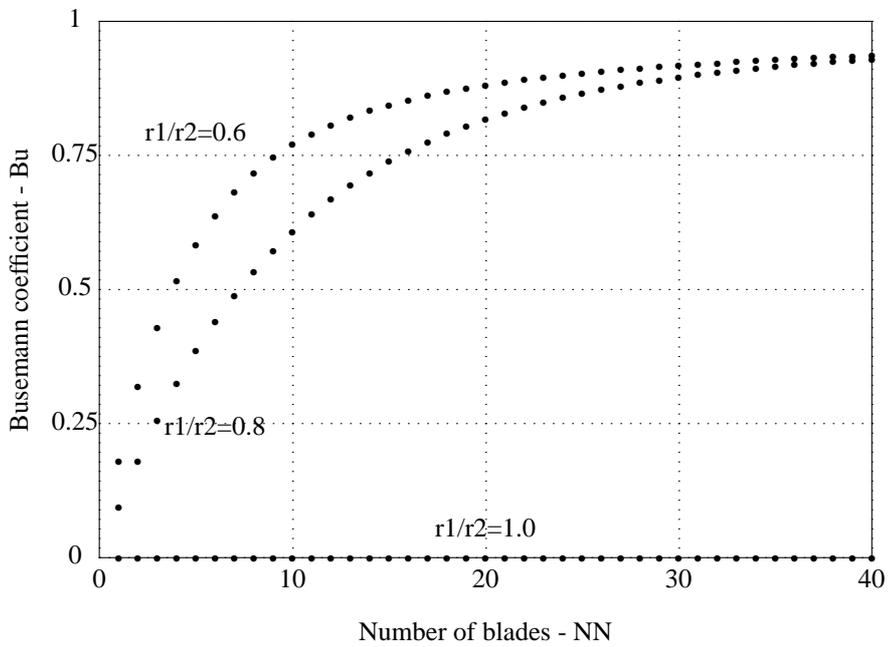

Fig 8. Thin straight blades $a = 1, 0 \leq q \leq 1, \beta = \pi/2$

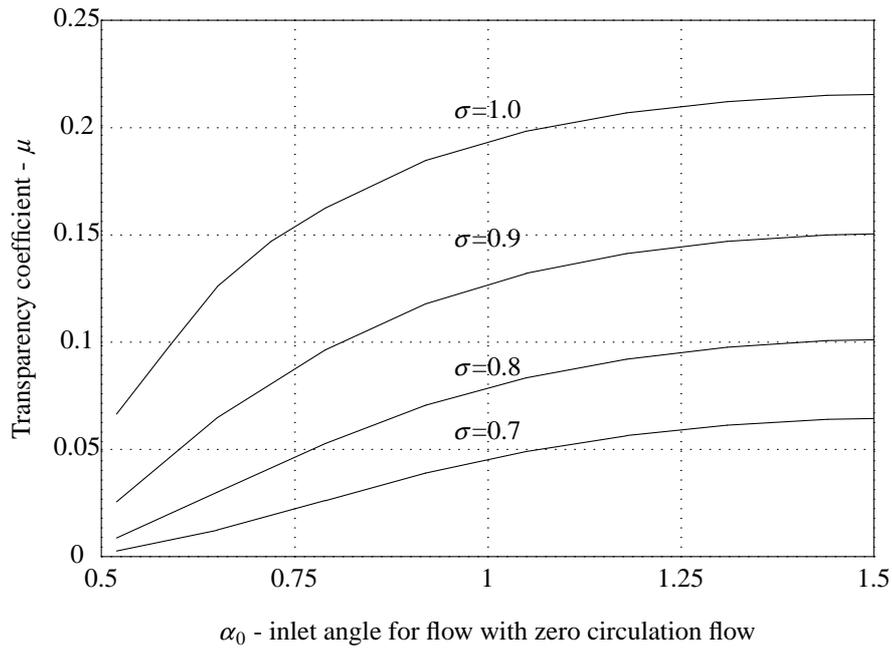

Fig 9. Common case N=6

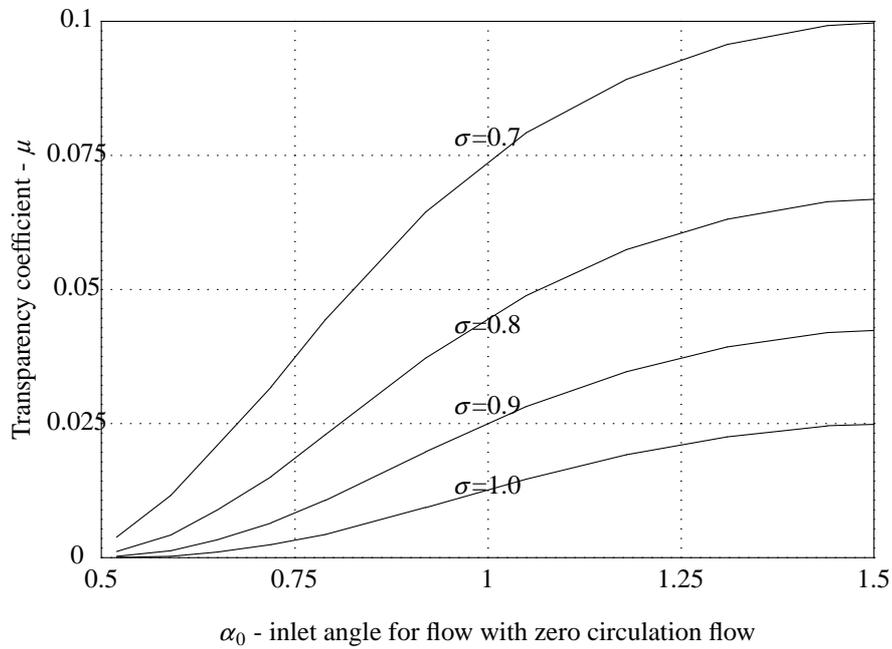

Fig 10. Common case N=9

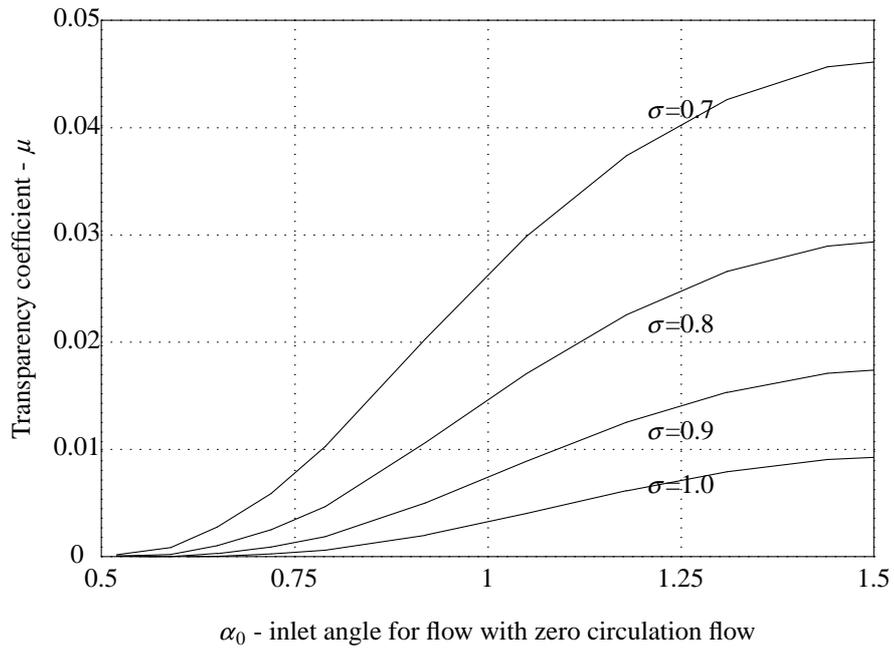

Fig 11. Common case N=12

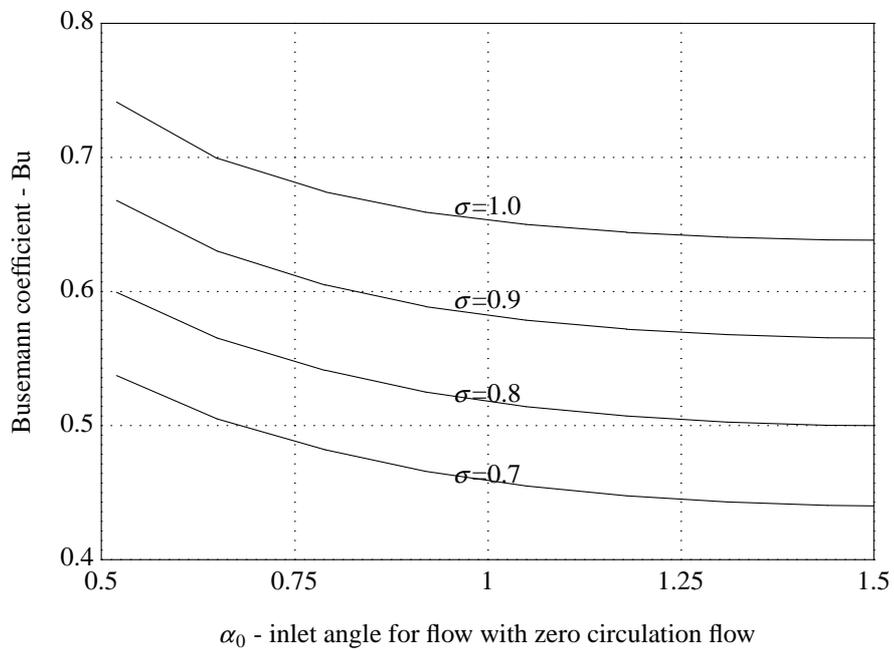

Fig 12. Common case N=6

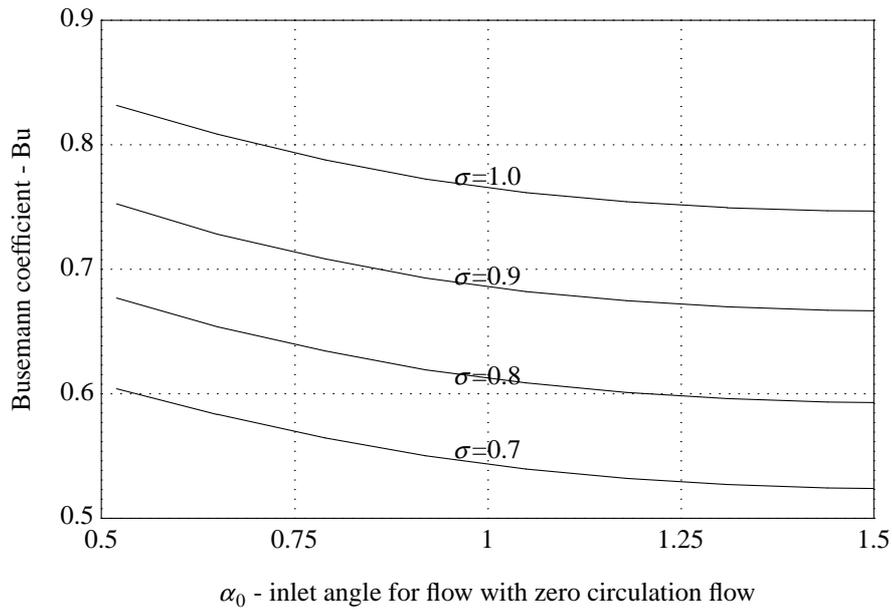

Fig 13. Common case N=9

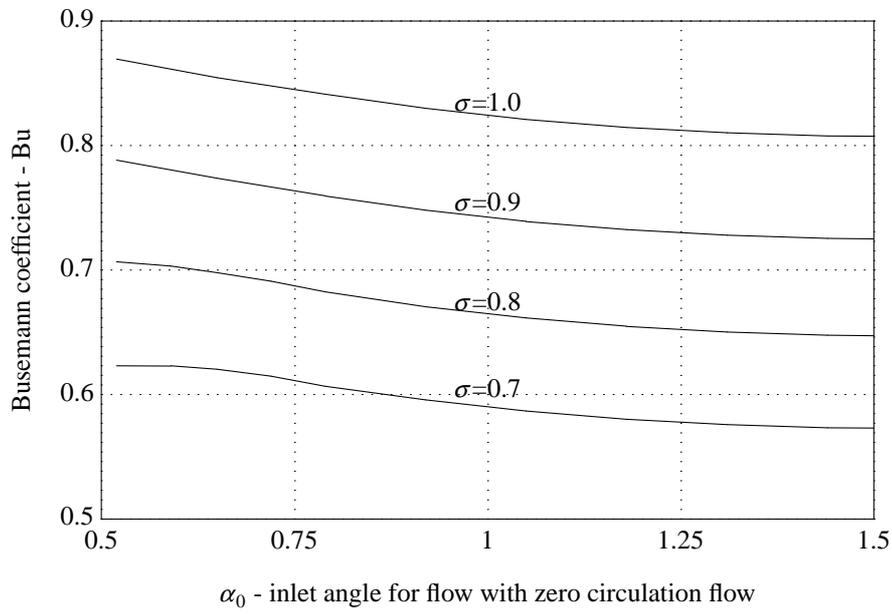

Fig 14. Common case N=12